\documentclass[10pt,conference]{IEEEtran}

\usepackage{amsmath, amssymb, amsbsy}
\usepackage{graphicx}
\usepackage{color}
\usepackage{hero}

\title{Routing in Mobile Ad-Hoc Networks using Social Tie Strengths and Mobility Plans}

\author{Riten Gupta\IEEEauthorrefmark{1}, 
Niyant Krishnamurthi\IEEEauthorrefmark{1}, 
Uen-Tao Wang\IEEEauthorrefmark{2}, 
Tejaswi Tamminedi\IEEEauthorrefmark{1}, 
and Mario Gerla\IEEEauthorrefmark{2} \\
\IEEEauthorrefmark{1}UtopiaCompression Corp., Los Angeles, CA\\
\{riten, niyant\}@utopiacompression.com\\
\IEEEauthorrefmark{2}UCLA, Los Angeles, CA
}
\begin{document}
\maketitle

\begin{abstract}
We consider the problem of routing in a mobile ad-hoc network (MANET) for which
the planned mobilities of the nodes are partially known a priori and the nodes
travel in groups. This situation arises commonly in military and emergency
response scenarios. Optimal routes are computed using the most reliable path
principle in which the negative logarithm of a node pair's adjacency
probability is used as a link weight metric. This probability is estimated
using the mobility plan as well as dynamic information captured by table
exchanges, including a measure of the social tie strength between nodes. The
latter information is useful when nodes deviate from their plans or when the
plans are inaccurate. We compare the proposed routing algorithm with the
commonly-used optimized link state routing (OLSR) protocol in ns-3
simulations. As the OLSR protocol does not exploit the mobility plans, it
relies on link state determination which suffers with increasing mobility. Our
simulations show considerably better throughput performance with the proposed
approach as compared with OLSR at the expense of increased overhead. However,
in the high-throughput regime, the proposed approach outperforms OLSR in terms
of both throughput and overhead.
\end{abstract}

\section{Introduction}

Mobile ad-hoc networks (MANETs) have been studied for several decades and, with
recent advances in wireless communication technology, are expected to find
increasing use in a wide range of applications. In general a MANET must be
designed to accommodate arbitrary mobility of the constituent nodes (under
reasonable speed constraints). The nodes may or may not move in groups, and
their trajectories are generally unknown. Many MANET routing algorithms and
protocols, both proactive and reactive, have been proposed over the years
\cite{pathan2009routing}, some of which take advantage of known characteristics
of the underlying network, and others which operate in the most general case.
For example, the LANMAR protocol \cite{pei2000lanmar} is specifically designed
for networks with group mobility, while Optimized Link State Routing (OLSR)
\cite{RFC3626},
makes no assumptions on mobility. Many more such examples of
environment-specific protocols exist. See \cite{pathan2009routing} for a
summary. In the last decade, researchers have shown that routing in
delay-tolerant networks (DTNs) can be aided by dynamically gathered information
about the social graph \cite{easley2010networks} of the underlying network
\cite{daly2007social} \cite{hui2011bubble}. In particular, a node's {\it
  centrality}, which in a social sense expresses its relative importance, can
be used to determine its suitability for forwarding packets.

In this paper, we examine MANETs formed by a set of nodes participating in a
collaborative activity for a finite time duration. An example is a military
setting in which the nodes consist of mounted and dismounted soldiers carrying
out a carefully planned mission. The mission plan includes the trajectory of
each node, specified by waypoints, and nodes travel in groups. The trajectory
information is useful for predicting node pair adjacencies and the group nature
of the node travel gives rise to a novel adjacency prediction method using
social tie strengths, which is beneficial when nodes deviate from their planned
trajectories.  This type of planned network, referred to as a {\it tactical
  edge} network, also arises in emergency response; for example, in disaster
relief missions. A network of unmanned aerial vehicles (UAVs) with known flight
plans is another potential example.  Previous studies \cite{lu2014information}
\cite{le2015social}
showed good results in using a ``social tie centrality'' metric for packet
forwarding in information-centric networks. In this work, we derive a novel
routing algorithm for tactical edge networks which exploits mobility plans and
disseminates social tie information 
in a manner similar to \cite{lu2014information}. 
Both pieces of information are used
to make routing decisions by the algorithm, which we call Tactical Edge Network
Social Routing (TENSR). We compare the performance of this algorithm in terms
of throughput, delay, and overhead, to OLSR for several representative cases
We find that significant throughput improvement relative to OLSR can be
obtained by exploiting the mobility plan and social information.

The remainder of this paper is organized as follows. In Section 
\ref{sec:networks} we introduce the tactical edge networks that are the
target applications for TENSR. The routing algorithm is then
described in detail in Section \ref{sec:routing}. We present ns-3 simulation
studies in Section \ref{sec:sim}. Finally, Section \ref{sec:conclusion}
concludes the paper.

\section{Tactical Edge Networks}
\label{sec:networks}

In this section we describe the characteristics of a tactical edge MANET both
qualitatively and mathematically. We also discuss how these features can be
used to aid routing decisions. We assume that $N$ network nodes carry out a
mission that spans $T$ seconds.

\subsection{Mobility Plans}
\label{sec:mobilityplans}

In a tactical edge MANET each node is associated with a mobility plan, which
consists of a finite set of waypoints describing the node's planned trajectory
throughout a mission. In most cases, the nodes travel in groups as they carry
out the mission. The nodes' actual trajectories are expected to deviate
somewhat from their mobility plans. In some cases, such as unplanned emergency
events, a node may drastically change its trajectory. In others cases
deviations will be much smaller and can be modeled by a random jitter
process. Mobility plans represent a very useful set of a priori information
that can facilitate routing decisions.

For node $i$, the mobility plan consists of a list of pairs 
$(t_1, \theta_i(t_1)), \dots, (t_n, \theta_i(t_n))$ where $n$ 
is the number of waypoints in node $i$'s
mobility plan, $t_j$ is the time of the $j$th waypoint, and $\theta_i(t_j) \in
\mathbb{R}^2$ is a 2-dimensional Euclidean position vector.
\footnote{We focus on 2-D mobility in this paper but point out that the 
analysis can be extended to the 3-D case in a straightforward manner.}
The list of waypoints is considered a discrete sampling of a node's trajectory.
Between waypoints, the node is assumed to travel at constant velocity along
the line segment connecting the two points. Thus, for each node a complete
planned trajectory exists: $\{\theta_i(t) : t \in [0,T] \}$.

\subsection{Position Location Information}

In military and emergency response networks, position location information
(PLI) may be available, for example from a blue force tracking system. 
Such a system communicates
position information about each node to all nodes. In this study, we assume
that PLI is disseminated on a channel independent of the network traffic.  
This information may be inaccurate (due to errors in the geolocation
mechanism) and unreliable (due to latencies or congestion in the dissemination
protocol). Let $\phi_i(t) \in \mathbb{R}^2$ be the PLI estimate of node $i$'s
location at time $t$.

\subsection{Social Tie in Tactical Edge Networks}
\label{sec:socialtie}


As we discuss in Section \ref{sec:routeselection}, in a MANET with uncertain
link state, we become interested in the probability that a pair of nodes is
adjacent--that is, whether a link between the nodes exists in the {\it
  communication graph}--for this information proves useful in route selection.
 If we define a {\it social graph} consisting of the
same vertices (nodes) as the communication graph, with edge weights
corresponding to the social tie strengths between pairs of nodes, then this
graph can be useful in estimating the probability of adjacency in the
communication graph.

Tie strengths and their implications for social networks were proposed in
\cite{granovetter1973strength} and extended in \cite{marsden1984measuring},
while the application of tie strengths to MANET routing is analyzed in
\cite{daly2009social}. Several factors are important in defining a social tie
metric, including intimacy/closeness, frequency of contacts, reciprocity, and
recency. In \cite{lu2014information}, a social tie metric based on frequency
and recency was shown to be effective in routing for delay-tolerant MANETs. In
our algorithm, we use a simple social tie definition which takes into account
only the frequency of encounters, as this lends itself to a simple adjacency
probability estimator. Specifically, we define the social tie $R_{i,j}(t)$
between nodes $i$ and $j$ at time $t$ to be the number of encounters between
these two nodes in the last $T_{mem}$ seconds. The parameter $T_{mem}$ is
called the {\it social tie memory}.  Here an encounter refers to the nodes
being adjacent at some point during a measurement interval
$T_{int}$\footnote{This definition of social tie is a special case of the one
  in \cite{lu2014information}, with $f(x)$ the indicator function of the
  interval $[0,T_{mem}]$.}.  Letting $R_{mem} = T_{mem} / T_{int}$ (assumed an
integer), we see that $R_{i,j}(t) \le R_{mem}$ and that the probability of
adjacency between nodes $i$ and $j$ can be estimated as
\begin{equation}
P(i,j \ \text{adjacent}) \approx \frac{R_{i,j}(t)}{R_{mem}}.
\label{eq:prob_adj_st}
\end{equation}

\section{Routing Algorithm}
\label{sec:routing}

Here we describe the routing algorithm. The algorithm consists of two main
activities: table exchanges and route selection.  The goal of the
table exchange process is to supply each node with a sufficent amount of
information from which it can estimate node adjacency probabilities.  These
probabilities are then used to determine most reliable paths. 

\subsection{Link State Data}
\label{sec:data}

We first describe the various data structures used by the nodes to exchange
information and estimate adjacency probabilities. Let $q$ be the index of 
an arbitrary node in the network.

\subsubsection{Social Tie}

Node $q$ stores an $N \times N$ matrix $\bR^{(q)}$ of estimated social ties
between each node pair.\footnote{All matrices described here are symmetric and
  only the upper or lower triangular parts must be stored}  Associated with
each entry in this matrix is a timestamp indicating when the value was
determined. These timestamps are stored in matrix $\bT^{(q)}_{ST}$. These
matrices are updated dynamically throughout the mission.

\subsubsection{Empirical Adjacency}

While the social tie measures the number of encounters, the empirical adjacency
of a pair of nodes indicates whether or not the nodes are currently adjacent.
These empirical adjacencies are stored in matrix $\bA^{(q)}$ and their
corresponding timestamps, indicating the time at which the adjacency (or
non-adjacency) was determined, are stored in matrix $\bT^{(q)}_{Adj}$. They
are updated dynamically

\subsubsection{Pairwise Distance}

Node $q$ has a pairwise-distance matrix $\bD^{(q)}$. The $(i,j)$th element
of this matrix is the measured geographic distance between nodes $i$ and $j$.
The corresponding timestamp matrix is $\bT^{(q)}_D$. These matrices are updated
dynamically during node encounters.

\subsubsection{Mobility Plans}

Node $q$ has the complete list of waypoints for each node. Thus, as described
in Section \ref{sec:mobilityplans}, node $q$ has knowledge of $\theta_i(t)$
for all $i \in \{1, \dots, N\}$ and $t \in [0,T]$. This information is known
a priori and is static (i.e., not updated dynamically).

\subsubsection{PLI}

At any time $t$, node $q$ has an estimate of each other node's (say node $i$'s)
position in the form of PLI $\phi^{(q)}_i(t)$. The accuracy of this information
depends in large part on its staleness. Thus node $q$ also stores the timestamp
for each other node's PLI: $T^{(q)}_{p,i}$.

\subsubsection{Other Parameters}

Several other parameters needed by the routing algorithm are described here.
The variance of each node's (say node $i$'s) trajectory $\sigma_{n,i}^2$
is known by all nodes. The variance of the PLI estimate is also known,
but since this depends on staleness, the nodes store a set of breakpoints
defining a piecewise linear function representing the PLI variance versus
the staleness, denoted $\sigma^2_{p}(\tau)$ where $\tau$ is the staleness.
Each node thus has knowledge of the function $\sigma_{p}^2(\tau)$. The maximum
communication range of each node $d_{\max,i}$ is known by all nodes, as is
the maximum velocity $v_{\max,i}$.

\subsection{Information Exchange}

Location information from a priori mission trajectories and PLI is an
invaluable aid for routing. However, nodes may deviate from their planned
trajectories and PLI may be inaccurate or stale. Thus, dynamic information is
also used to make routing decisions. Specifically, a node pair's adjacency,
distance, and social tie are used. In order for nodes to get this information,
a dissemination method is employed consisting of handshaking and table
exchanges. Although the purpose of this dissemination method is to give the
nodes the information necessary for routing, it must also minimize
communications overhead to be useful. In this section, we briefly describe this
information dissemination process. Please refer to \cite{wang2016} for more
detail. 

Information exchanges are done using two types of messages: HELLO messages
and INFO messages. A HELLO message is a small message sent periodically
which is used to identify neighbors. A HELLO message also conveys small
pieces of information to the recipient. An INFO message is used to transmit
table information from one node to another. This information is either
adjacency, distance, or social tie. The average time interval between
emissions of HELLO messages by a node is called the {\it HELLO interval}. This 
interval is a design parameter. A low HELLO interval results in greater
information sharing, leading to better route selection and improved
throughput. However, this also entails a larger communication overhead as
many HELLO messages are transmitted per unit of time. Similarly, the
{\it INFO interval} is the minimum time duration between emissions of INFO
messages by a node. This interval is also a design parameter. The 
tradeoffs described for the HELLO interval apply to the INFO interval as well.

Table exchanges are done during node encounters. During an encounter, a pair of
nodes determines the geographic distance between the pair. It is assumed that
the nodes have the necessary processing hardware to make this estimate. The
nodes also update their social tie and adjacency information and decide what
information to transmit to one another. In \cite{wang2016}, the table exchange
process, along with a description of the message formats, is given. Note that
prior knowledge of social ties can be used to initialize the nodes' social
tie matrices, but in general the social structure is considered a dynamic
process which is tracked throughout the mission.

\subsection{Route Selection}
\label{sec:routeselection}

Since in a tactical edge network, link adjacencies are usually not known with 
certainty, routes may not be chosen based on typical shortest path principles.
Instead, we choose paths that are \textit{most reliable} \cite{guerin1999qos}.
Let $\calS_{m,n}$ be the set of all paths between nodes $m$ and $n$ in the 
complete graph. Each such path $s$ is a set of links which may be represented
as node pairs. So for example, if the link between nodes $i$ and $j$ is in path
$s$ then $(i,j) \in s$. The most reliable path between $m$ and $n$ is
\begin{eqnarray}
s_{m,n}^* &=& \underset{s \in \calS_{m,n}}{\argmax} 
\prod_{(i,j) \in s} \text{Pr} (i,j \text{ adjacent}) \nonumber\\
&\approx& \underset{s \in \calS_{m,n}}{\argmax} 
\prod_{(i,j) \in s} \hat{p}_{i,j} \nonumber\\
&=& \underset{s \in \calS_{m,n}}{\argmin} 
\sum_{(i,j) \in s} -\log \hat{p}_{i,j} 
\label{eq:mrp}
\end{eqnarray}
where $\hat{p}_{i,j}$ is an estimate of $ \text{Pr} (i,j \text{ adjacent})$.
Here we are assuming that the link adjacency probabilities are independent.
From (\ref{eq:mrp}), optimal routes can be found using Dijkstra's algorithm
on a graph with edge weights given by $-\log \hat{p}_{i,j}$ for edge $(i,j)$.

\subsection{Link Adjacency Probability Estimation}

Most reliable path (MRP) route selection requires that each node has an
estimate of adjacency probability for each node pair in the network. The
adjacency probabilities can be estimated in a variety of ways, depending on
what information is available to the estimating node. These methods are
described in this section.  In the following we assume that node $q$ is
estimating the probability of adjacency of nodes $i$ and $j$, with $i \ne j$,
(but $q$ may or may not equal $i$ or $j$).

\subsubsection{Case 1: $q \in \{i,j\}$}

If $q$ is one of the nodes in the pair for which adjacency probability is
needed, then the most accurate estimate can be found using node $q$'s adjacency
matrix. The handshaking mechanism ensures that each node always knows its
neighbor list accurately \cite{wang2016}. Therefore, we set 
$\hat{p}_{i,j} = \bA^{(q)}_{i,j}$.

\subsubsection{Case 2: Using Previous Adjacency Information}

Assuming $q \notin \{i,j\}$, in some cases the adjacency of nodes
$i$ and $j$ can still be confirmed with certainty. Suppose $\bA^{(q)}_{i,j}=1$
and let $\tau = t - \bT^{(q)}_{D,i,j}$ where $t$ is the current time. Thus
$\tau$ is the staleness of the pairwise distance estimate $d =\bD^{(q)}_{i,j}$.
Then if
\begin{equation*}
\tau \le \tau_{thresh} = \frac{d_{\max} - d}{v_{\max,i} + v_{\max,j}}  
\end{equation*}
(where $d_{\max} = \min(d_{max,i}, d_{\max,j})$), then the nodes must currently
be adjacent. Therefore, if $\tau \le \tau_{thresh}$, then we set
$\hat{p}_{i,j} = 1$.

\subsubsection{Case 3: Using Location Information}

Both a priori trajectory plans and PLI help to determine adjacencies. Actual
trajectories differ from the waypoint specification due to (small) jitter
processes that occur even when the nodes follow their mission plans, but also
due to emergencies and changes of plans which cause large differences with
respect to the waypoints. PLI is known to be noisy due to latency and error
in geolocation.

\subsubsection*{Gaussian Models}

We model the deviation between trajectories and trajectory plans (based on
waypoints) as Gaussian with known covariance $\sigma_{n,i}^2 I$. Similarly, the
PLI error is modeled as Gaussian with staleness-dependent covariance
$\sigma_{p,i}^2(\tau) \cdot I$.  In general, it is expected that PLI error
variances are larger than a priori error variances. In the case of nodes that
deviate from the mission plan due to emergencies or other unplanned events, the
a priori trajectory provides no information regarding the true location. Our
approach is to first determine if a node has deviated significantly from its
plan using a simple hypothesis test based on observation of PLI. If such a
deviation has not occurred, then the node's location $X_i(t)$ is modeled as
Gaussian with mean equal to the planned location $\theta_i(t)$ and covariance
$\sigma_{n,i}^2$, or $X_i(t) \sim \calN(\theta_i(t), \sigma_{n,i}^2 \cdot I)$.
If a deviation has occurred, then the location is modeled as Gaussian with mean
equal to the PLI $\phi_i(t)$, and covariance $\sigma_{p,i}^2(\tau) \cdot I$, or
$X_i(t) \sim \calN(\phi_i(t), \sigma_{p,i}^2(\tau) \cdot I)$.

\subsubsection*{Hypothesis Test}

Dispensing with time dependencies in the notation for now, we model the PLI as
$\phi_i = X_i + M_i$ where \mbox{$M_i \sim \calN(0, \sigma_{p,i}^2 \cdot
  I)$}. Next, let $\calH_0$ be the hypothesis that node $i$ is travelling
according to its plan, or
\begin{equation*}
\calH_0: X_i = \theta_i + N_i
\end{equation*}
with $N_i \sim \calN(0, \sigma_{n,i}^2 \cdot I)$.
Then $\calH_0$ can be written 
\begin{equation*}
\calH_0: \phi_i \sim \calN(\theta_i, \sigma^2 I)
\end{equation*}
where $\sigma^2 = \sigma_{p,i}^2 + \sigma_{n,i}^2$. Next, note that
\begin{equation*}
\frac{\| \phi_i - \theta_i \|^2}{\sigma^2} \sim \chi_2^2.
\end{equation*}
%
Let $1 - \alpha$ be a confidence level, where $0 < \alpha \ll 1$ and set
$\text{Pr}(\|\phi_i - \theta_i\| \le T | \calH_0) = 1- \alpha$,with $T$ a
threshold. Solving for $T$, $\calH_0$ should be rejected whenever 
\begin{equation*}
\| \phi_i - \theta_i \| > 
\sigma \sqrt{ F_{\chi_2^2}^{-1} (1 - \alpha) }
\end{equation*}
%
%
where $F_{\chi_2^2}$ is the cumulative distribution function (CDF) of a 
chi square variate with two degrees of freedom.

\subsubsection*{Location-Based Adjacency Probability Estimation}

The hypothesis test is conducted for both node $i$ and $j$. For node $i$, if
$\calH_0$ is rejected, then the location is modeled as 
\begin{equation*}
X_i(t) \sim \calN(\phi_i(t), \sigma_{p,i}^2(\tau) \cdot I).
\end{equation*}
Otherwise it is modeled as
\begin{equation*}
X_i(t) \sim \calN(\theta_i(t), \sigma_{n,i}^2 \cdot I).
\end{equation*}
For node $j$, a
similar model is obtained for $X_j(t)$. Thus let $(\mu_i, \sigma_i^2 I)$ and
$(\mu_j, \sigma_j^2 I)$ be the means and covariances of nodes $i$ and $j$
respectively. Then 
\begin{equation*}
\frac{\| X_i(t) - X_j(t) \|^2}{\sigma_i^2 + \sigma_j^2} \sim \chi_2^2(\lambda).
\end{equation*}
The non-centrality parameter of this chi-square random variable is
$\lambda = \| \mu_i - \mu_j \|^2 / (\sigma_i^2 + \sigma_j^2)$ and the 
probability of adjacency of nodes $i$ and $j$ is
\begin{equation*}
\hat{p}_{i,j} = 
\text{Pr}(\| X_i(t) - X_j(t) \| \le d_{\max}) = 
F_{\chi_2^2(\lambda)} \left( \frac{d_{\max}^2}{\sigma_i^2 + \sigma_j^2} \right)
\end{equation*}
where $F_{\chi_2^2(\lambda)}$ is the non-central chi-square CDF with non-centrality
parameter $\lambda$.

\subsubsection{Case 4: Using Social Tie}

Next, if PLI information is unavailable for one or both nodes $i,j$
(i.e., it is so stale that we consider it useless), then we use the
social tie between the nodes to estimate the adjacency probability.
Recall from equation (\ref{eq:prob_adj_st}) that the adjacency probability
estimate is 
\begin{equation*}
\hat{p}_{i,j} = \frac{R_{i,j}(t)}{R_{mem}}.  
\end{equation*}
This equation is used when the staleness of either node's PLI is greater
than a threshold, called the {\it PLI staleness threshold}.

\subsubsection{Case 5: Default Case} 

Finally, if the social tie is also deemed too stale to be useful, a default
non-zero adjacency probability is assigned: $\hat{p}_{i,j} = p_0$. Setting
$p_0$ to 0 would preclude the consideration of link $(i,j)$ in the MRP routing
process. On the other hand, setting $p_0$ too high will cause MRP to favor the
link, which may not exist. Thus a small positive value is preferred for $p_0$.
This value is used when the social tie staleness of nodes $i,j$ is greater
than a threshold, called the {\it social tie staleness threshold}.

\section{Simulation Results}
\label{sec:sim}

In this section, we present the results of several simulations comparing the
TENSR routing protocol to OLSR for various tactical edge networks. The
simulations are done using ns-3 \cite{henderson2008network}. The nodes follow
mobility plans specified by waypoints. These waypoints are known to the routing
protocol (they are part of the a priori information). However, only the
waypoints corresponding to the first half of the mission duration, i.e.,
$[0,T/2]$, are known. Beyond this, the mobility is unknown to the routing
protocol.

\subsection{Models and Parameters}

\subsubsection{Node Mobility}
\label{sec:mobility}

Nodes travel at constant velocity between waypoints, which are generated
randomly based on desired node velocities. A Gaussian jitter process with
covariance $\sigma_n^2 \cdot I$ is added to the linear movement. Nodes travel
in groups and all nodes within a group share the same mobility plan
(waypoints), but have independent jitters. We consider the following groupings:
7 groups with 3 nodes per group, 5 groups with 4 nodes per group, and a 10
groups of 2 nodes.  A geographical area with dimensions $1500 \times 1500$
square meters is considered.  Typically, tactical edge networks are connected;
the missions are designed as such. Therefore, when generating waypoints
randomly, we ensure a high likelihood of connectedness, or equivalently a low
average network partitioning (ANP) \cite{kurkowski2006two}.

\subsubsection{PLI Model}
\label{sec:pli}

PLI comes from a channel independent of the communication network; for example,
from a blue force tracking system. The PLI is sporadic, and not all nodes in
the network obtain the PLI at every PLI transmission. PLI is implemented in
simulation as an oracle residing on each node, which informs a random subset of
all other nodes of its position (with an additive noise component). The
following parameters characterize the PLI model. First the {\it broadcast
  interval} is the time interval at which PLI broadcasts are
sent. Additionally, a {\it broadcast interval jitter} is added to this interval
randomly. The {\it broadcast reach probability} is the probability of
successful PLI reception by any node. Finally the {\it broadcast delay} is the
delay of the PLI broadcast signal. 

\subsubsection{Traffic Types}

We simulated constant bit rate (CBR) traffic and used
a simple echo application in which a source node sends a packet to a
destination, which then replies to the source, only if and when it receives the
original packet. Six CBR flows are used with packet size 1024 bits and rate 1
packet/sec. The sources and destinations are always chosen to lie in separate
groups.

\subsubsection{Summary of Parameters}

\begin{table}[!t]
\caption{Simulation parameters}
\centering
\begin{tabular}{l|l}
\hline
Parameter & Values simulated \\
\hline\hline
Groups, nodes/group & (2,10), (3,7), (4,5) \\
\hline
Simulation duration, $T$ & 10 min \\
\hline
Deviation time ($T/2$) & 5 min \\
\hline
Node velocity & 10 m/s, 20 m/s, 30 m/s \\
\hline
Jitter standard deviation, $\sigma_n$ & 10 m (lat and lon) \\
\hline
Radio range & 500 m \\
\hline
Geographic area & $1500 \times 1500$ square meters\\
\hline
\# Monte-Carlo trials & 30\\
\hline
HELLO interval & 0.5 sec \\
\hline
INFO interval & 4 sec \\
\hline
Social tie memory, $R_{mem}$ & 10 \\
\hline
Social tie staleness threshold & 60 sec \\
\hline
PLI broadcast interval & 5 sec \\
\hline
PLI broadcast interval jitter & 5 sec \\
\hline
PLI broadcast reach probability & 0.5 \\
\hline
PLI broadcast delay & 3 sec \\
\hline
PLI standard deviation, $\sigma_p^2(\tau)$ & 10 m (lat and lon) at $\tau=0$\\
\hline
& 20 m (lat and lon) at $\tau=20$ sec\\
\hline
& 30 m (lat and lon) at $\tau=30$ sec\\
\hline
PLI staleness threshold & 10 sec \\
\hline
Hypothesis test confidence level & 95\%\\
\hline
OLSR HELLO interval & 2 sec \\
\hline
OLSR TC interval & 5 sec\\
\hline
\end{tabular}
\label{table:params}
\end{table}

Table \ref{table:params} summarizes the simulation parameters.

\subsection{Comparison with OLSR}

We first compare the TENSR protocol with OLSR in its standard configuration.
The standard OLSR configuration uses a HELLO interval of 2 sec and a topology
control (TC) interval of 5 sec \cite{RFC3626}. As these intervals are somewhat
analagous to TENSR's HELLO interval and INFO interval, and the TENSR intervals
are shorter than their OLSR counterparts, the results are expected to
favor TENSR. We will see that this is the case. Later, we discuss simulations
in which both protocols' intervals are approximately the same.

\begin{figure}[!t]
\centering
\includegraphics[width=3.5in]
{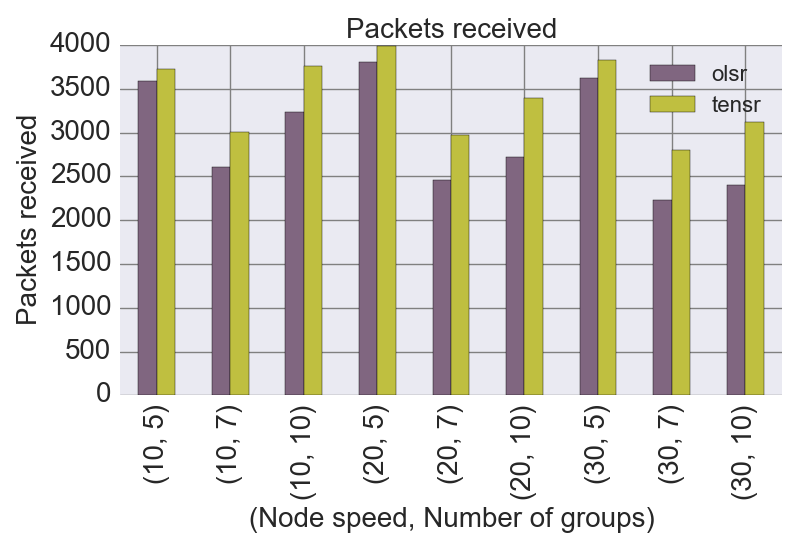}
\caption{Average number of packets successfully received with TENSR and OLSR}
\label{fig:cbr_pli50_h05_i8_BarChart_of_packets_receivedTime_of_Dev300}
\end{figure}

Figure \ref{fig:cbr_pli50_h05_i8_BarChart_of_packets_receivedTime_of_Dev300}
shows the average number of received packets over the course of 10 min for both
protocols for various node velocities and various node groupings with CBR
traffic. (See Table \ref{table:params}.)  We found that 30 Monte-Carlo trials
were enough to sufficiently estimate the mean.  The results correspond to the
following TENSR parameters: HELLO inerval = 0.5 sec, INFO interval = 4 sec.
The simulation duration is 10 min, and at 5 min, the nodes deviate from their
planned mobility pattern. In all cases, the TENSR throughput is superior to
that of (standard) OLSR.

\begin{figure}[!t]
\centering
\includegraphics[width=3.5in]
{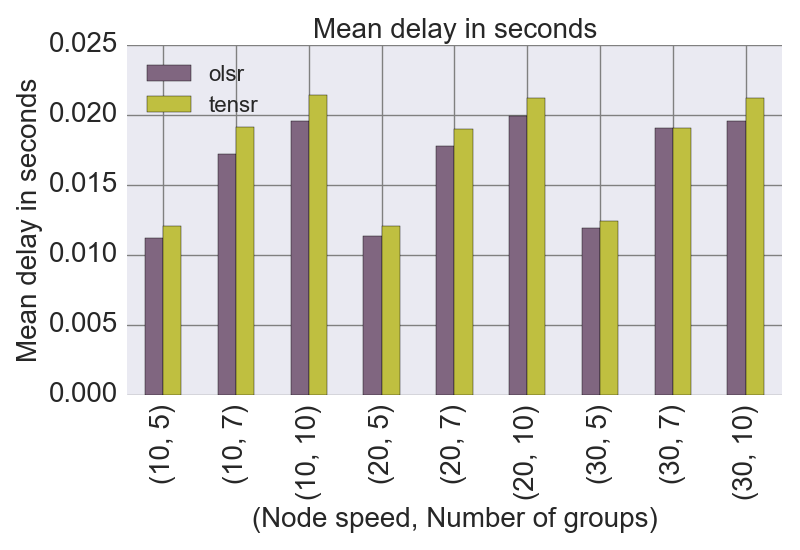}
\caption{Average packet delay with TENSR and OLSR}
\label{fig:cbr_pli50_h05_i8_BarChart_of_mean_delayTime_of_Dev300}
\end{figure}

Next, Figure \ref{fig:cbr_pli50_h05_i8_BarChart_of_mean_delayTime_of_Dev300}
shows the average packet delay for the two protocols, as well as over all
packets successfully received) for CBR traffic. Note that the TENSR delay is
higher than the corresponding OLSR delays. While this is true in an average
sense, a closer inspection shows that in fact packets that are successfully
transmitted in both cases (TENSR and OLSR) receive approximately the same
delay, whereas packets that are successfully transmitted by TENSR, but not OLSR
receive a higher delay. Since these packets do not impact the OLSR average
delay, the TENSR average delay is higher. Thus, more packets are transmitted
when using TENSR, but these ``extra'' packets experience a higher-than-average
delay. Other packets experience approximately the same delay with both
protocols.

\subsection{Throughput vs. Overhead Comparison}

\begin{figure}[!t]
\centering
\includegraphics[width=3.5in]
{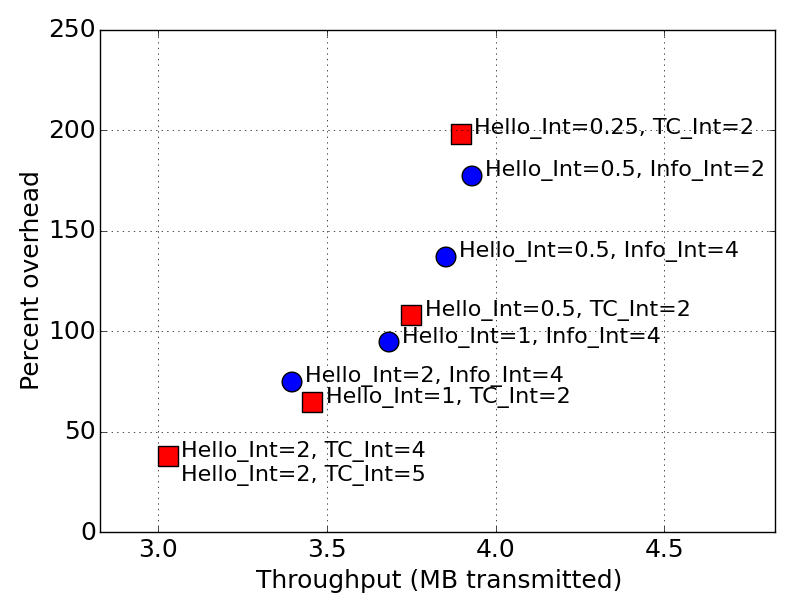}
\caption{Throughput vs. overhead. Circles are TENSR simulations, squares
are OLSR simulations.}
\label{fig:thpt_ovhd}
\end{figure}

The results of the previous section show that the proposed routing protocol can
exploit the mission and social information to deliver performance superior to
OLSR. However, this performance improvement comes at the expense of increased
communication overhead in the form of table exchange packets. The performance
of OLSR can be improved by changing its HELLO interval and topology control
interval \cite{huang2006tuning}.  In this section, we focus on a single
network configuration and run simulations with varying OLSR parameters. We also
run TENSR simulations with varying parameters. We use an expanded HELLO message
format \cite[Sec 3.4.3]{wang2016} which includes some link-state information in
the HELLO message and which generally improves performance.  We log the number
of overhead bits and compare the percent overhead versus throughput for both
protocols. These simulations use the following parameters: 3 groups, 7 nodes
per group, 10 min simulation time, without deviation from planned mobility, and
20 m/s velocity, The HELLO interval, INFO interval, OLSR hello interval, and
OLSR TC interval are varied. The remaining parameters are identical to those in
Table \ref{table:params}.

Figure \ref{fig:thpt_ovhd} shows the throughput versus percent overhead for
both OLSR and the TENSR for various parameter sets. Generally, the
two protocols are comparable and tend to lie on the same curve. 
As the HELLO, INFO, and TC intervals become small, the throughput
reaches a limit. In this regime, the proposed protocol provides slightly higher
throughput than OLSR for less overhead (approximately 10\%). Thus, we conclude
that while the mission and social information is certainly helpful for routing,
in some cases, its benefits can be gained by using OLSR with more messaging
(and overhead). In the high-throughput regime, however, the proposed protocol
is superior to OLSR, even with increased OLSR messaging.

\section{Conclusions}
\label{sec:conclusion}

We have developed a novel routing protocol for tactical edge networks which
exploits known mission information such as mobility plans, as well as dynamic
social information and position location information. The protocol has been
compared to OLSR and shown to be superior in terms of throughput and
approximately equal in terms of delay, at the expense of increased overhead.
In the high-throughput regime, which is achived by reduction of the HELLO
and information exchange intervals in both protocols, the TENSR protocol
is superior to OLSR in both throughput and overhead.

\section{Acknowledgements}

The authors wish to thank Dr. You Lu from UCLA and Dr. Jacob Yadegar from
UtopiaCompression Corporation for helpful insights.  This material is based
upon work supported by the RDECOM-CERDEC-STG-TW under Contract
No. W56KGU-15-C-0069.

\bibliography{IEEEabrv,tensr}
\bibliographystyle{IEEEtran}

\end{document}